\documentclass{optica-article}

\journal{oe}

\articletype{Research Article}

\usepackage{lineno}
\usepackage{xcolor}
\usepackage{ulem}

\newcommand{\eref}[1]{Eq.~(\ref{#1})}
\newcommand{\sref}[1]{Sec.~\ref{#1}}
\newcommand{\fref}[1]{Fig.~\ref{#1}}

\begin{document}

\title{High-efficiency coupled-cavity optical frequency comb generation}

\author{M. P. Mrozowski,\authormark{1,*} J. Jeffers,\authormark{1} and J. D. Pritchard\authormark{1}}

\address{\authormark{1} Department of Physics, University of Strathclyde, SUPA, Glasgow G4 0NG, UK}


\email{\authormark{*}mateusz.mrozowski@strath.ac.uk} 



\begin{abstract}\label{abstract}
We present a high efficiency source of picosecond pulses derived from a dual cavity optical frequency comb generator. This approach overcomes the limitations of single cavity comb generators that are restricted to efficiencies of a few percent. We achieve picosecond pulses with GHz repetition rates offering over a hundred times higher output efficiency than a single cavity design and demonstrate tuning of pulse width by varying the modulation depth of the intra-cavity electro-optic modulator. 
These results provide a wavelength-agnostic design with a compact footprint for the development of portable picosecond pulsed laser systems for timing, metrology, and LIDAR applications.
\end{abstract}
    
\section{Introduction}\label{sec:0}
Optical frequency comb technology underpins a wide range of applications in sensing and metrology \cite{Udem:99,Telle2001,Fortier19}. As well as enabling precision measurements, frequency combs also provide an important resource for generating large numbers of temporal and frequency modes for realising scalable quantum systems \cite{kues19}. Typically these experiments utilise combs derived from bulky, high power and optically pumped mode-locked laser systems generating temporal pulses with widths ranging from a few fs to 100s of ps 
and a 
typical repetition rate ($\sim$ 1 - 100 MHz). These mode-locked laser systems additionally find applications in quantum LIDAR via efficient generation of heralded single photons, for example in a birefringent single mode optical fiber \cite{Smith:09,England_2019}.

One route towards compact, efficient frequency combs has been enabled by recent advances in cavity-based electro-optic frequency combs in thin-film lithium niobate\cite{Hu_2022} utilising the general critical coupling scheme proposed in\cite{Hu_2021}. While these devices offer very large free spectral ranges and a compact footprint \cite{Pasquazi18,doi:10.1126/science.aan8083}, the inherent material losses in waveguide technologies limit the wavelengths over which these devices can be utilised, as well as the maximum theoretical efficiency achievable and, furthermore, they cannot be dynamically tuned.

An alternative approach to generating optical frequency combs is to use an electro-optic modulator (EOM) \cite{Parriaux:20}. An optical frequency comb generator (OFCG) using an intra-cavity EOM can convert a CW pump into a mode-locked pulse train\cite{Kobayashi72,kourogi,Kovacich:00,Xiao:09}. These simple devices are able to produce optical pulses with extremely high repetition rates up to 20 GHz, whilst achieving pulse widths of order hundreds of ps down to a few hundred femtoseconds\cite{Xiao:09}. 
Advantages of this approach are the design can be adapted to operate at arbitrary pump wavelength using appropriate optics, the pulse width can be dynamically adjusted by changing the EOM modulation depth, as well as a very low timing jitter that makes it suitable for the distribution of precision timing signals \cite{Xiao:08}. 
As an example of OFCG versatility, the device can be combined with spectral pulse shaping to realise a ps-pulse arbitrary waveform generator in the telecom band \cite{Jiang07}. 
The main drawback of the single cavity OFCG is the intrinsically low output efficiency (typically $<1\%$) which arises from the imperfect impedance matching when coupling light into the OFCG \cite{Kobayashi72}.

In this paper we present compact and wavelength agnostic source of few-ps optical pulses based on a dual-cavity OFCG\cite{Macfarlane:96} that overcomes the efficiency limitation of the single cavity design. 
We demonstrate an output efficiency exceeding 80\%, achieving a minimum pulse width of 1.77~ps, in excellent agreement with a parameter-free model of the coupled cavity system. 
This provides a simple, low-power source suitable for applications in quantum LIDAR \cite{mrozowski}.

\section{Coupled Cavity OFCG}\label{sec:1}
\begin{figure}[t!]
\centering\includegraphics[width=\textwidth]{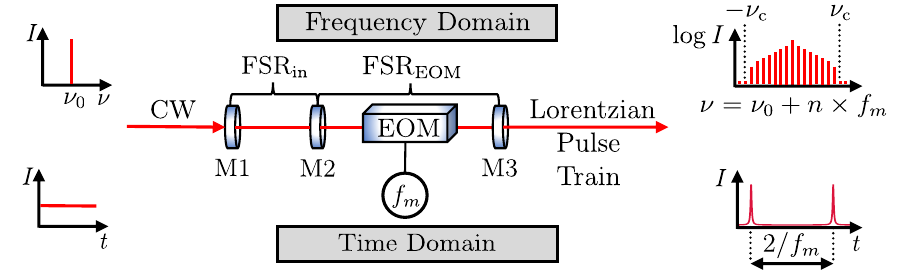}
\caption{ Dual-cavity OFCG formed from a CW seed laser injected into an impedance matched input cavity (M1 \& M2) with free spectral range $\mathrm{FSR_{in}}$ to pump the EOM cavity (M2 \& M3) with free spectral range $\mathrm{FSR_{EOM}}$ chosen to be an integer multiple of $=f_m$, the EOM modulation frequency. This configuration generates a high efficiency optical frequency comb, with short temporal pulses at a repetition rate of $2f_m$ and a broad frequency spectrum showing a sharp cut-off at $\pm\nu_c$, the characteristic mode resonant with the input cavity.}
\label{Fig:1a}
\end{figure} 
 
A single cavity OFCG source uses an EOM inside an optical cavity driven resonantly at the cavity free spectral range (FSR) to generate a comb of phase-coherent sidebands that convert a CW pump laser into a train of short optical pulses that have a Lorentzian line shape and a repetition rate equal to twice the modulation frequency \cite{doi:10.1063/1.1654403,Kovacich:00,kourogi}. 
The output efficiency of the single cavity OFCG is \cite{Xiao:08}  
\begin{equation}
\label{eq:1}
\eta_{\mathrm{SC}} = \frac{(1-\mathrm{R})^2}{\pi\mathrm{R}(1-\mathrm{R}\eta_c)\beta},
\end{equation}
where $\mathrm{R}$ is the power reflectivity of the resonant cavity mirrors, $\eta_c$ is the crystal transmission efficiency and $\beta$ is the modulation depth in radians. 
The pulse width at FWHM intensity is given by \cite{doi:10.1063/1.1654403}
\begin{equation}
\label{eq:2}
\Delta\tau_p = \frac{1}{2F\beta f_m},
\end{equation}
where $F$ is the finesse of the resonant cavity and $f_m$ is the modulation frequency of the EOM. Equations (\ref{eq:1}) and (\ref{eq:2}) show that it is not possible to generate short optical pulses with high output efficiency, as increasing either cavity finesse or modulation depth increases the loss from the central cavity mode, resulting in an impedance mismatch that inhibits efficient input coupling into the cavity and hence low output, with <1$\%$ being typical for experiments \cite{Kovacich:00,Xiao:09}.

This limitation can be overcome using a coupled cavity configuration, as first demonstrated in \cite{Macfarlane:96} and shown schematically in \fref{Fig:1a}. Here the system consists of three mirrors M1-3, with power reflectivities $\mathrm{R_{1-3}}$ forming two optical cavities, the input cavity (M1-M2) and the EOM cavity (M2-M3) containing the modulator crystal. 
In this configuration, the presence of the input cavity with $\mathrm{R}_1=\mathrm{R}_2$ allows for critical impedance matching of the OFCG system, dramatically increasing the output power and theoretically enabling 100\% input coupling into the EOM cavity formed by M2-M3.

In our setup we drive the EOM crystal at a modulation frequency $f_m$ chosen to be an integer multiple $n$ of $\mathrm{FSR_{EOM}}$, the FSR of the EOM cavity. For the coupled cavity setup, an additional restriction on the frequency width of the generated comb arises when the $n^{\mathrm{th}}$ sideband of the EOM becomes resonant with the $j^{\mathrm{th}}$ mode of the input cavity 
\begin{equation}
\label{eq:3}
n \times f_m = j \times \mathrm{FSR_{in}}.
\end{equation}
To maximise the comb-width (and hence minimise pulse width), $\mathrm{FSR_{in}}$ is chosen to have a large common denominator with $f_m$ and a large finesse to reduce the input cavity linewidth. The first sideband that is completely resonant with the input cavity is called the characteristic mode $\nu_c$, which defines the maximum sideband order populated in the comb.

The added complexity of the dual-cavity operation and characteristic mode causes the efficiency and pulse-width to deviate from the simple formulae presented above for the single cavity case. Instead it is necessary to solve numerically the detailed model presented in Ref. \cite{mrozowski} describing the interplay of the two cavities. 
Experimental parameters were chosen by using this model to find the required mirror reflectivities and cavity FSRs to give high efficiency pulses with a target duration of 1.7~ps and at a 4.8~GHz repetition rate. In the figures below we also use this model for theoretical comparison to experimental data. Furthermore, it is essential to ensure that the mirror radii of curvature are chosen to give a matching waist for the fundamental modes of both cavities in order to maintain spatial mode-matching conditions and suppress excitation of transverse modes of the EOM cavity.

\section{Experiment Setup}\label{sec:2}
\begin{figure}[t!]
\centering\includegraphics[width=\textwidth]{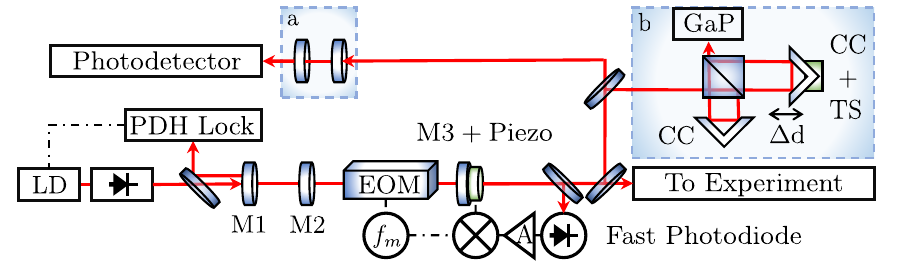}
\caption{ Schematic of the experimental setup showing a CW laser diode ($\mathrm{LD}$) with a tandem isolator locked to the input cavity using a Pound-Drever-Hall lock. 
The EOM cavity length is stabilised using a piezo actuator on M3 locked using an error signal derived from demodulation of a fast-photodiode signal. 
Frequency and time domain of the output pulses are analysed in
a : A Fabry-Perot analysis cavity. 
b: A auto-correlation setup with time variable delay, here CC is a corner cube, TS is a motorised translation stage used to generate time delay $\mathrm{2\Delta d}$/c and GaP is a slow GaP photodetector.}
\label{Fig:1b}
\end{figure} 

The experimental setup for our realisation of the dual-cavity OFCG is shown schematically in \fref{Fig:1b}. The seed light for the system was provided by a distributed bragg reflector  laser diode at 780 nm, with a maximum power output to the experiment of 30 mW. Due to strong optical feedback from the input cavity a 60 dB tandem isolator is used, and the diode laser is fiber coupled into a single mode polarisation maintaining fiber to allow efficient spatial mode-matching into the input cavity.

The input cavity (M1-M2) is a plano-concave cavity with $R_1=R_2=0.99$ formed using an input radius of curvature of 1000~mm and a separation of 12.6~mm tuned to obtain a measured FSR of $\mathrm{FSR_{in}}$ = 11.85 $\pm$ 0.05 GHz. 
The cavity is constructed using a monolithic brass mount to minimise sensitivity to vibration. Mirror M2 is oriented with the antireflection (AR) coated surface located inside the input cavity to supress losses within the EOM cavity, resulting in a measured input cavity finesse of $280\pm10$. 
The seed light is stabilised to the input cavity using a Pound-Drever-Hall lock, using current modulation to add sidebands to the laser output at 4~MHz. 
We obtain 95\% transmission of light into the EOM cavity when the laser is locked to the input cavity limited by the spatial mode-matching of light into the cavity.

The intracavity EOM is chosen to have a resonant frequency of $f_m = 2.39$~GHz. The EOM crystal has length $\mathrm{L_c} = 27$ mm, group velocity dispersion GVD = 447 fs$^2$/mm and transmission efficiency $\eta_c$ = 0.996. In the experiments below we tune the modulation depth $\beta$ by adjusting the applied microwave power to give $\beta$ in the range of 0.3 to 2.05~rad, actively stabilising the EOM resonance frequency using feedback to a peltier cooler that controls the crystal temperature. 

The EOM cavity is formed by adding mirror M3 with reflectivity $R_3=0.96$ and radius of curvature of 300~mm, chosen to mode-match the input cavity waist of 150~$\mu$m. 
The finesse of the EOM cavity is restricted to 100 due to the losses caused by the finite crystal transmission and the imperfect AR coatings of the crystal facet. Following initial alignment, the EOM cavity length
is adjusted to obtain a free spectral range $\mathrm{FSR_{EOM}}$ = 478 MHz, corresponding to the fifth subharmonic of the modulation frequency $f_m$. 
In addition to matching the FSR by adjusting cavity length, the position of the EOM crystal must also be adjusted to ensure the crystal is located at an anti-node of the microwave modulation frequency to ensure the relative phase of successive cavity round-trips add constructively. 
The EOM crystal is initially coarsely placed at $\sim\lambda_m$ from M2, and then its location is adjusted using a translation stage to maximise the suppression of the fundamental mode to optimise the round-trip RF phase matching condition \cite{Bell:95}.
Here we use the fact that the power in the central frequency decreases as the modulation depth increases.
The EOM cavity is stabilised by using a piezo actuator on M3. A dispersive error signal is obtained by picking off a fraction of the light transmitted through M3 and directing it to a fast photodiode. 
The fast photodiode signal is then demodulated at the modulation frequency $f_m$ and low-pass filtered. We utilise the fact that at the exact resonance the power of the $f_m$ harmonic almost vanishes \cite{Xiao:09}. 
The error signal is then amplified and a fast Red-Pitaya based PID controller\cite{redpitaya} is used to apply feedback to the piezo.

\section{Experimental Results}\label{sec:3}
\begin{figure}[t!]
\centering\includegraphics[width=.75\textwidth]{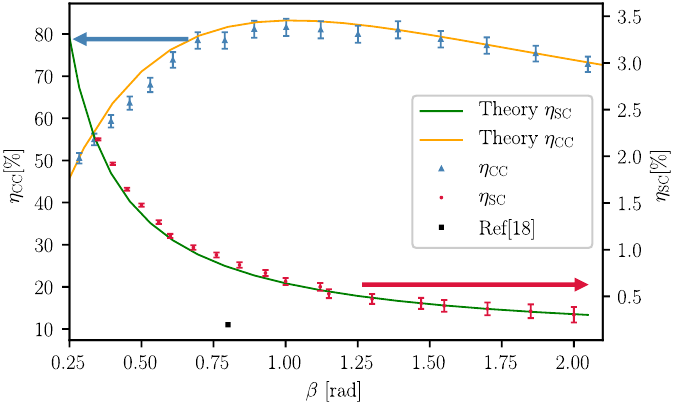}
\caption{Output OFCG efficiency for the single cavity OFCG - $\mathrm{\eta_{SC}}$ on the right-hand $y$-axis and the coupled cavity OFCG - $\mathrm{\eta_{CC}}$, on the left-hand $y$-axis. The black square data-point is efficiency recorded in \cite{Macfarlane:96}.}
\label{Fig:2}
\end{figure} 

To evaluate the performance of the coupled cavity OFCG, we compare the output efficiency as a function of modulation depth against that obtained in the single cavity configuration with mirror M1 removed. For the single cavity data the OFCG cavity is locked as described above with the input laser free-running. 
Results are shown in \fref{Fig:2}, with errorbars representing the standard deviation of 30 measurements at each modulation depth. 
The maximum output efficiency for the coupled cavity configuration is $81\pm 2$\% at $\beta=1~$rad, more than a 100$\times$ improvement compared to the single cavity configuration, and $7\times$ higher than the previous demonstration of a coupled cavity setup \cite{Macfarlane:96}. 

The single cavity OFCG efficiency data $\eta_{\mathrm{SC}}$ are compared to parameter free theoretical curve taken from \eref{eq:1}, whilst the theoretical coupled cavity efficiency $\eta_{\mathrm{CC}}$ was generated using the numerical model from Ref.~\cite{mrozowski}, using parameters given in \sref{sec:2}. We obtain good agreement with theory for the high values of $\beta$ > 1.25 but a slight deviation for $\beta < 0.75$. This can be attributed to the significantly lower bandwidth of the error signal when $\beta$ is reduced in both single and coupled cavity OFCG configurations. In the case of the coupled cavity system, additional errors occur at low modulation depth due to the increased leakage of fundamental mode seed light from the EOM cavity into the input cavity which perturbs its lock, preventing stable operation at lower values of $\beta$.   

In order to characterise the pulse train output from the dual cavity OFCG, some of the output light is picked off and directed to an analysis cavity and an autocorrelation module in order to characterise the frequency and temporal properties as shown in Fig.~\ref{Fig:2}. The analysis cavity consists of two mirrors with identical power reflectivity $\mathrm{R_a}$ = 0.99 and an FSR of $\mathrm{FSR_a}$ = 130 GHz. The autocorrelation module is a scanning Michelson interferometer using travelling corner cubes to adjust relative pulse delay between the two arms \cite{Wong:94}. The interferometer output is detected using a slow GaP photodiode and a high-gain, low noise current amplifier.

\begin{figure}[t!]
\centering\includegraphics[width=\textwidth]{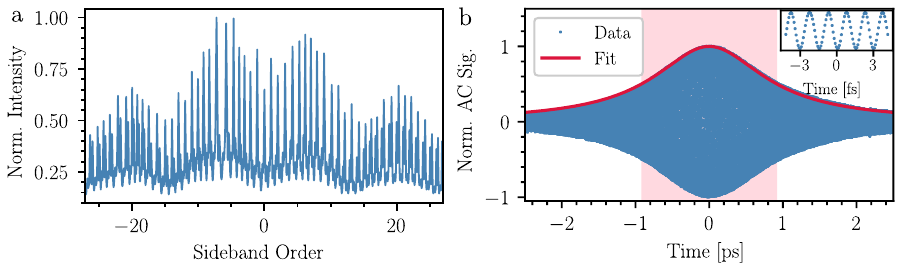}
\caption{a: OFCG Frequency spectrum measured using an analysis cavity with $\mathrm{FSR_a}$=130~GHz. Additional modes seen on the plot can be attributed to the imperfect transverse mode suppression in the analysis cavity, and to frequency combs overlapping due to $\mathrm{FSR_a}$ being narrower than the produced frequency comb. b: The normalised field autocorrelation signal measured using a scanning Michelson interformeter. The solid red line is a Lorentzian fit function used to determine the FWHM pulse width $\Delta\tau_{p}$\ = \ 1.77 $\pm$ 0.04 ps. The inset plot shows the field autocorrelation signal on a femtosecond scale. Both traces were obtained using $\beta$\ =\ 2.05 rad.}
\label{Fig:3}
\end{figure}

The frequency spectrum for the dual-cavity OFCG operating at $\beta=2.05$~rad is shown in  \fref{Fig:3}a, corresponding to a FWHM bandwidth of $\sim$120 GHz. Using the fact the OFCG output pulses should be Lorentzian, this corresponds to a transform-limited FWHM pulse width of $\sim$1.75 ps. Here the spectral envelope shows a deviation from the monotonic exponential decay with sideband order expected for a Lorentzian, however calculating the Fourier transform of the observed spectrum with the appropriate phase relationship caused by the round-trip phase accumulated in the EOM cavity shows excellent agreement with a Lorentzian pulse shape. This deviation in spectral output is likely due to residual amplitude modulation, which has been observed in previous experiments\cite{doi:10.1063/1.1654403,Brothers:97,477736}. Comparing the single and dual cavity output spectra we observe an identical envelope verifying this arises from the EOM cavity.

In order to measure the pulse width independently we record the field autocorrelation as shown in \fref{Fig:3}b, using the observed interference fringes to calibrate pulse delay. Due to the relatively low peak pulse power we are unable to observe a two-photon absorption feature using the GaP detector, and this data therefore corresponds to the linear field autocorrelation signal \cite{Wong:94}. The temporal envelope is fit using a Lorentzian to obtain a FWHM pulse width of $1.77\pm0.04$~ps, in good agreement with the spectral characterisation using the analysis cavity. 

\begin{figure}[t!]
\centering\includegraphics[width=.5\textwidth]{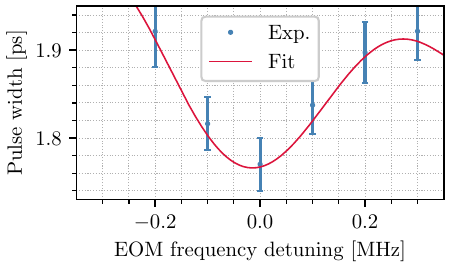}
\caption{The FWHM pulse width as a function of the EOM cavity frequency detuning. For each EOM resonance frequency the cavity length is adjusted to ensure the cavity FSR as a harmonic of the modulation frequency. Solid line shows a fit to the theoretical model of Ref.~\cite{mrozowski}. Data presented here are obtained with a modulation depth of $\beta$ = 2.05 rad.}
\label{Fig:4}
\end{figure}

The OFCG output pulse width depends strongly on the characteristic mode $\nu_c$ of the EOM cavity that becomes resonant with a mode of the input cavity. The monolithic design of the input cavity prevents scanning of $\mathrm{FSR_{in}}$.  Instead we scan the FSR of the EOM cavity by adjusting the frequency of modulation $f_m$ and temperature-tuning the EOM back into resonance, followed by adjusting the physical EOM cavity length to ensure it remains an integer multiple of the modulation frequency. This approach means we can scan FSR$_\mathrm{EOM}$ without introducing a relative detuning between the EOM and the cavity, such that changes in pulse width arise only from a change in the ratio of FSR$_\mathrm{EOM}$ to FSR$_\mathrm{in}$.
  
Results are shown in \fref{Fig:4}, where the effect of changing the ratio between $\mathrm{FSR_{in}}$ and $f_m$ on the measured pulse width can be observed corresponding to an increase in pulse width as the characteristic mode number is reduced  as the EOM cavity is tuned away from the optimum operating point. We additionally add the predictions of the theoretical model, which shows excellent agreement with the observed changes.

\begin{figure}[t]
\centering\includegraphics[width=.75\textwidth]{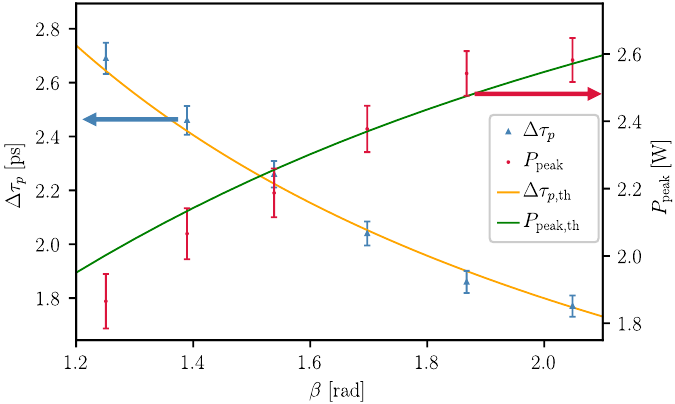}
\caption{The FWHM pulse width and peak power per pulse as a function of the modulation depth $\beta$. $P_{\mathrm{peak}}$ was recorded for input optical power equal to 30 mW.}
\label{Fig:5}
\end{figure}

Finally, we measure pulse width and hence extract peak power as a function of the modulation depth. Results are shown in \fref{Fig:5}, with the drop in output efficiency for $\beta>1$ observed in \fref{Fig:3} compensated by the reduction in pulse width $\propto1/\beta$ corresponding to a peak pulse power of 2.6 W. Again we see good agreement with the theoretical model of Ref. \cite{mrozowski}, with a larger deviation in the predicted peak power at lower modulation depths primarily due to the reduction in input cavity lock stability. These results imply higher power would be possible using increased modulation depth, however for $\beta>\pi$ the output pulse spacing becomes unstable.

\section{Conclusion}\label{sec:4}
In conclusion, we present a highly efficient, high repetition rate source of pulsed light. With an input power of 30~mW, peak powers of $\sim$ 2.6~W were observed at a repetition rate of 4.78~GHz, achieving pulse widths down to 1.77~ps limited by the maximum accessible modulation depth of the EOM. Our results show excellent agreement with a parameter-free theoretical model for the OFCG performance \cite{mrozowski}, and  demonstrate the ability to realise an OFCG with over 80\% output efficiency.
To obtain higher output power, we have repeated the characterisation using a high power, narrow linewidth Ti:Sapph laser enabling use of up to 300~mW into the OFCG. Our measurements show the OFCG retains the same high efficiency and pulse width observed with low power laser, however even for this increased peak power of up to 26~W we still are unable to observe a two-photon intensity autocorrelation spectrum using GaP detector. The current limitation in operating with higher input powers arises from the damage threshold of the AR coating on the EOM of 20~W/mm$^2$. This, combined with the relatively small active crystal aperture of $2\times2$~mm required to maintain a GHz resonance frequency limits us to 300~mW seed. In future, this can be overcome using alternative coatings or EOM crystal materials, or for example operating at telecom wavelengths to benefit from higher damage thresholds.
Already the output pulse power demonstrated here with the dual cavity OFCG is sufficient to enable heralded photon pair generation in a birefringent optical fiber \cite{mrozowski}, and this design paves the way to realising compact, low cost and high repetition rate picosecond pulses for a wide range of potential applications in metrology, sensing and quantum lidar.

\begin{backmatter}
\bmsection{Acknowledgements}\label{sec:acknowledgements}
We thank E. Riis for useful discussions and loan of equipment. This project is funded by the UK Ministry of Defence.
\bmsection{Disclosures}
The authors declare no conflicts of interest.
\bmsection{Data availability}\label{sec:dataavailability}
The datasets used in this work are available at \cite{data}.
\end{backmatter}



\bibliography{sample}






\end{document}